# Interplay of spin-precession and higher harmonics in the parameter estimation of binary black holes

N. V. Krishnendu[*] and Frank Ohme[†]

*Max Planck Institute for Gravitational Physics (Albert Einstein Institute),
Callinstr. 38, D-30167 Hannover, Germany
and Leibniz Universitat Hannover, D-30167 Hannover, Germany*

Gravitational-wave signals from coalescing compact binaries carry an enormous amount of information about the source dynamics and are an excellent tool to probe unknown astrophysics and fundamental physics. Though the updated catalog of compact binary signals reports evidence for slowly-spinning systems and unequal mass binaries, the data so far cannot provide convincing proof of strongly precessing binaries. Here, we use the gravitational-wave inference library parallel Bilby to compare the performance of two waveform models for understanding the spin-induced orbital precession effects in simulated binary black hole signals. One of the waveform models incorporates both spin-precession effects and subdominant harmonics. The other model accounts for precession but only includes the leading harmonic at quadrupolar order. By simulating signals with varying mass ratios and spins, we find that the waveform model with subdominant harmonics enables us to infer the presence of precession in most cases accurately. On the other hand, the dominant harmonic model often fails to extract enough information to measure precession. In particular, it cannot distinguish a face-on highly-precessing binary from a slowly-precessing binary system irrespective of the binary's mass ratio. As expected, we see a significant improvement in characterizing precession for edge-on binaries. Other intrinsic parameters also become better constrained, indicating that precession effects help break the correlations between mass and spin parameters. In contrast, spin-precession measurements are prior dominated for equal-mass binaries with face-on orientation, even if we employ a waveform model including subdominant harmonics. In this case, doubling the signal-to-noise ratio does not help to reduce these prior induced biases. As we expect detections of highly-spinning binary signals with misaligned spin orientations in the future, simulation studies like ours are crucial for understanding the prospects and limitations of gravitational-wave parameter inferences.

## I. INTRODUCTION

The LIGO Scientific Collaboration and the Virgo Scientific Collaboration have released an updated catalog of gravitational-wave (GW) detections, GWTC-3, containing about ninety GW events [1–17]. This new set of events include many exceptional candidates such as GW190814 [10], GW190412 [9], and GW190521 [11] and allows us to perform qualitatively new studies of astrophysical populations and fundamental physics [14,18–21]. In the future more of these detections are predicted [22] following upgrades of the detector sensitivities. The upcoming decade is going to provide us a wide variety of GWs from compact binary mergers. Future third-generation detectors [23–27] and space-based detectors [28–31] may also be operational by then.

Once we have detected GW signals, we must analyze them to infer the source properties. Individual masses, spin magnitudes, angles specifying the orientation of spins, the location, and the orientation of the source, etc., characterize a generic binary system. One of the main challenges here is making correct and accurate measurements of these binary parameters, employing faithful waveform models and efficient algorithms.

The observed GW data to date provides ample evidence of unequal mass binary systems and hence opportunities to probe higher signal harmonics beyond the dominant quadrupole. During the third observing run (O3) of LIGO and Virgo, a binary system was detected with the primary black hole (BH) being ∼3.75 times more massive than the secondary; GW1901412 [9]. Only a few months later, LIGO and Virgo announced the observation of GW190814 [10], in which the mass asymmetry is even larger (∼8.9) and the nature of the secondary is widely debated in the literature [32–37]. We emphasize that the current template-based binary search pipelines do not include higher harmonics and spin-precession effects, but the recent developments may circumvent such selection biases in the future [38–42].

[*]krnava@aei.mpg.de
[†]frank.ohme@aei.mpg.de

The confident detections of higher harmonics in these systems provide more information about the source properties, mainly through breaking the degeneracy between the luminosity distance and inclination angle [43]. Further, they enable us to perform tests of general relativity on a completely different source population and expand our knowledge about the astrophysics of such binaries [17,43–49].

Though the present data is insufficient to provide strong constraints on the individual spin magnitudes and their respective orientations in most binaries, it is possible to make statements about the statistical evidence for the binary system's aligned-spin and in-plane spin components. The updated catalog of compact binary signals reports evidence for aligned-spin components and moderate spin-induced orbital precession. Noticeably, the two candidates GW190412 [9] and GW190521 [11,50,51] show intriguing hints for a nonzero value of spin-induced orbital precession. We also note that the orbital eccentricity signatures of the heaviest binary black hole (BBH) system GW190521 are extensively studied in the literature [52–54].

Knowing whether the source is precessing or not is very important from an astrophysical and fundamental physics point of view [43,55–79]. Notably, the spin orientation measurements have immediate implications for determining the binary formation channels [43,64,80–85], distinguishing between binaries formed in isolation [86–97], through dynamical evolution [98–110], or specific scenarios like hierarchical formation [111–116]. Moreover, numerical relativity simulations of BBHs show a significant dependence of the remnant's kick velocity on the BH spin orientations [117–124]. However, spin precession is inherently difficult to measure, especially as most sources are detected with small inclination angles [125–135]. In addition, several studies have suggested that inferring the precession spin parameter is difficult as it is often prior dominated, and systematic waveform differences might be significant enough to bias the result in many cases [136–143].

Here we investigate how including subdominant harmonics may alleviate some of the problems. Focusing on precessing binary BH systems with and without mass asymmetries, we explore the possibility of characterizing spin-induced orbital precession effects varying the source inclination angle to the detector. We compare the performance of two different waveform models highlighting the importance of using higher harmonics for accurately measuring the spin-precession effects. We also investigate the impact of two different signal-to-noise ratios to quantify the possible improvements to perform such measurements once we have the future GW detector facilities operational.

Gravitational waveform templates play an important role in GW searches [41,79,144–147] as well as for inferring the correct binary source properties [148–153]. The effect of subdominant harmonics in the GW parameter estimation is studied in [149] using a complete Bayesian analysis for a three-detector network. This study considered nonprecessing binaries of total mass ∼120 $M_\odot$ (source frame) and different mass ratios using the NR-surrogate model `NRHybSur3dq8` [154]. The main finding of that study is that the exclusion of higher harmonics in the parameter estimation analysis induces systematic biases for non-precessing BH binaries. Binaries with spins antialigned to the orbital angular momentum tend to provide more biased results than those with aligned spins, especially for signals with moderate signal-to-noise ratios. This inference is insufficient to conclude the possible systematic biases for binaries with spin-induced orbital precession.

Similarly, the GW parameter estimation study performed in Ref. [155] focuses on the importance of higher harmonics in analyzing nonprecessing binary BHs. For binaries with a total mass of ∼100 $M_\odot$ and varying mass ratios $q = m_2/m_1 = 0.5, 0.25, 0.12$, it is found that the estimates are largely biased for edge-on systems with significant mass asymmetry.

In Ref. [148], the measurability of spin-precession effects for GW190814-like and GW190412-like systems was explored in detail. The authors found that the spin-precession effects are measurable with reasonably good accuracy for asymmetric systems with moderate precession. Also, they demonstrated that a relatively small amount of precession can lead to a systematic offset in the inferred binary parameters. This offset in the parameter recovery may arise either from the difference in the signal model (`SEOBNRv4PHM` [156]) and the templates (`IMRPhenomPv2` [157] and `IMRPhenomD` [158,159]) used to analyze the simulated injections, or it can arise from other effects such as prior/orientation induced biases that propagate from extrinsic to intrinsic parameters.

In our analysis, we focus on the spin-induced orbital precession parameter measurements for generic binary systems using waveform models that account for both spin-induced orbital precession effects as well as higher harmonics using a complete Bayesian analysis. For most of our study, we consider the same waveform model for describing the signal and the template manifold. This aids us to disentangle the true systematics mainly coming from the unmodeled effects present in the template manifold and the prior induced biases.

We note that the Bayes factor would be the appropriate quantity to assess whether the data contain statistically significant traces of precession. However, we focus on measuring the effective precession parameter instead of the Bayes factor for two reasons. First, accurately estimating the Bayes factors would require calculating the evidence for precessing and nonprecessing binaries. Here we only perform fully precessing analyses. While it is possible to complement our study with the appropriate nonprecessing analyses for comparison, deriving robust Bayes factors has, in our experience, proved sensitive to analysis choices that have nothing to do with the effect we want to quantify. This is mainly due to the significant increase in the parameter-space dimensions when including precession. Therefore, we choose to only report Bayes factors for higher-harmonics vs dominant-harmonics models that are based on sampling identical parameter spaces.

Second, we want to avoid entering discussions of what we consider a precessing binary. Generically, all the BH spins in nature are expected to be misaligned with the orbital angular momentum and each other, if only by a minuscule amount. Therefore, only significantly strong precession would constitute a meaningful measurement of precession. To define what that means, we would have to impose limits on a continuous parameter (such as the misalignment angle or the precession parameter). While such an approach is valid and statistically robust, we prefer focusing directly on the measurement of a continuous parameter that carries information about the strength of precession.

This paper is organized in the following way. Section II describes the theoretical foundations of decomposing the GW signal into harmonics and the relation to characterizing the spin precession effects in generic binary system. Section III introduces our method. The main findings of this analysis are given in Sec. IV. We conclude in Sec. V.

## II. WAVEFORM DECOMPOSITIONS

Before presenting our main analysis of injected signals using complex parameter-estimation techniques, we first introduce the basic structure of GW signals. In particular, we summarize three signal decompositions below: 1. The angular dependency of the signal in the inertial frame can be described efficiently in terms of spin-weighted spherical harmonics. 2. Spin-induced orbital precession introduces amplitude and phase modulations in the inertial spherical harmonics that arise through a time-dependent rotation of the coprecessing spherical harmonics. 3. Precessing signals can be expanded in terms of the angle between the total and orbital angular momentum. Unambiguously identifying precession requires measuring at least two terms in this expansion. The relation between these decompositions helps to build an expectation for how subdominant harmonics affect measuring precession.

For a generic binary system, the total angular momentum $\vec{J}$ is the vector sum of its orbital ($\vec{L}$) and the individual spin-angular momenta ($\vec{S}_1$ and $\vec{S}_2$),

$$\vec{J} = \vec{L} + \vec{S}_1 + \vec{S}_2. \quad (2.1)$$

We choose a frame in which $\vec{J}$ is along the $z$-axis. In the case of simple precession, the direction of $\vec{J}$ changes much less than the directions of $\vec{L}$ and $\vec{S}_i$, and one can treat $\vec{J}/\|\vec{J}\|$ to a good approximation as constant [160]. See Fig. 1 for a visual representation of the coordinate system and for the angles that we define in it.

In this frame, we decompose the complex gravitational waveform $h = h_+ - ih_\times$ in terms of the spin weight-2 spherical harmonics,

$$h(t, \vec{\lambda}, \theta_{\rm JN}, \phi) = \sum_{\ell \geq 2} \sum_{-\ell \leq m \leq \ell} h_{\ell, m}(t, \vec{\lambda})_{-2} Y_{\ell m}(\theta_{\rm JN}, \phi). \quad (2.2)$$

Here, $\vec{\lambda}$ represents the set of intrinsic parameters, and $(\theta_{\rm JN}, \phi)$ are the spherical angles that define the direction of the line-of-sight. $_{-2}Y_{\ell m}$ are the spherical harmonic functions and $h_{\ell,m}$ are the spherical harmonic modes, which we refer to as *inertial* spherical harmonics to stress the frame of decomposition.

Previous research found that spin precession introduces phase and amplitude modulations in the $h_{\ell m}$ [160]. However, if the decomposition is carried out in a coprecessing frame that follows the instantaneous movement of the orbital plane, those modulations disappear [161]. In fact, Schmidt *et al.* [162] identified the coprecessing harmonics as the ones of a nonprecessing binary with the same masses and spins defined by the projection of the fully precessing spins onto $\vec{L}$. Reference [157] used this to construct the inertial harmonics of precessing systems via the rotation

$$h_{\ell,m} = \sum_{-\ell \leq m' \leq \ell} h_{\ell,m'}^{\rm np} D_{m',m}^{\ell}(\alpha, \beta, \epsilon), \quad (2.3)$$

where $D_{m',m}^\ell(\alpha, \beta, \epsilon)$ is the Wigner-$D$ matrix that depends on the angles $(\alpha, \beta, \epsilon)$ describing the instantaneous orientation of the orbital plane [163].[1] See Fig. 1 for details about the various angles. The superscript "np" denotes the harmonics of the associated nonprecessing system. The Wigner-$D$ matrix can be further expanded as

$$D_{m',m}^\ell(\alpha, \beta, \epsilon) = e^{im\alpha} d_{m',m}^\ell(-\beta) e^{-im'\epsilon}, \quad (2.4)$$

where $d_{m',m}^\ell$ is the small Wigner matrix that only depends on the opening angle $\beta$ between the total and angular momentum. The relation in Eq. (2.3) is only approximately valid. It neglects modifications to the remnant black-hole ringdown and asymmetries between the $(\ell, m)$ and $(\ell, -m)$ modes that are present in precessing systems [164].

We stress that this rotation mixes simple coprecessing harmonics ($h_{\ell,m}^{\rm np}$) into all inertial harmonics ($h_{\ell,m}$) of a precessing system that have the same $\ell$. Some of the first precessing waveform models only employed the dominant ($\ell = 2, |m| = 2$) nonprecessing harmonics [156,157,165]. Therefore, the precessing signals included only the inertial ($\ell = 2, |m| \in \{1, 2\}$) harmonics. $h_{2,0}$ is typically neglected because both the coprecessing $h_{2,0}^{\rm np}$ is very small, and $d_{2,0}^{\ell} \sim \sin^2(\beta/2)$ is small for moderate opening angles $\beta$.

---

[1]$\epsilon$ is defined in terms of $\alpha$ and $\beta$ as $\dot{\epsilon} = \dot{\alpha}\cos\beta$, where over dot represents the time derivative.

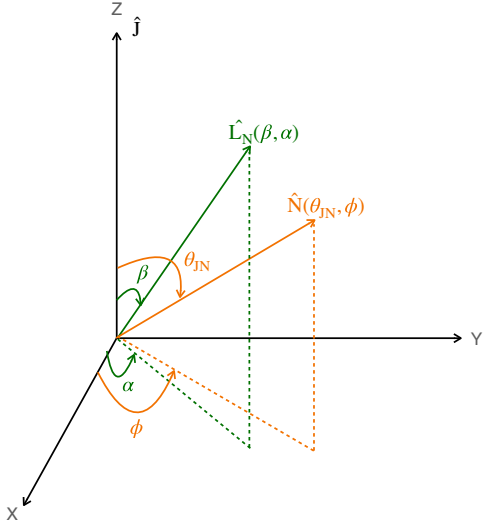

FIG. 1. Binary system whose total angular moment vector ($\hat{J}$) points to the z-axis and the orbital angular momentum and line-of-sight directions described by ($\beta, \alpha$) and ($\theta_{\rm JN}, \phi$) respectively.

This angle $\beta$ characterizes the amount of precession. The precession time scale is equivalent to the time-variation of the three precession angles and is much longer than the orbital time scale of the binary, but much shorter than the radiation reaction time scale during the inspiral. Hence, precession manifests itself as overall modifications of the $h_{\ell, m}$. However, for a single mode, these modifications could be mimicked by biased intrinsic parameters and cannot be identified unambiguously as the effect of precession.

To address the problem of detecting precession effects in GW signals, a harmonic decomposition for spin-precession signals has been proposed in Refs. [166–168]. They found that the leading-order signal (containing only harmonics with $\ell = 2$) can be decomposed into five *precession* harmonics using the expansion parameter $b = \tan \beta/2$. The authors then conclude that, in order to identify precession effects in the signal unambiguously, one requires simultaneous measurements of two *precession* harmonics.

The *precession* harmonics are not identical to the mode decomposition shown in Eq. (2.2) but are a linear combination of a subset of the spherical harmonics provided at specific orientations (fixed values of $\theta_{\rm JN}$ and $\phi$). Hence, the argument in Refs. [166,167] about spin-precession measurements may be translated to a more general notion of mode decompositions; i.e., one needs two harmonics with different $\beta$ dependencies to accurately identify precession effects. As we shall illustrate next, those two terms could come from two harmonics in the inertial frame or two coprecessing harmonics that get mixed into one inertial-frame harmonic.

Let us consider two examples to understand the precession effects on waveform models either including higher coprecessing harmonics or only including dominant harmonics. First, for a waveform model that only includes the ($\ell = |m| = 2$) coprecessing harmonics, we can use Eq. (2.3) to express the inertial harmonics of the precessing model as

$$h_{2,2} = \sum_{m'=\pm 2} h_{2,m'}^{\rm np} e^{i2\alpha} d_{m',2}^2(-\beta) e^{-im'\epsilon} \quad (2.5)$$

and

$$h_{2,1} = \sum_{m'=\pm 2} h_{2,m'}^{\rm np} e^{i\alpha} d_{m',1}^2(-\beta) e^{-im'\epsilon}. \quad (2.6)$$

Even though there are no contributions from higher coprecessing harmonics, the contributions from $h_{2,1}$ can become significant for highly-precessing binaries, where $d_{m',1}^2(-\beta)$ is nonzero. In order to measure $h_{2,1}$, we require $_{-2}Y_{21}(\theta_{\rm JN}, \phi) \neq 0$ as evident from Eq. (2.2). If the binary is oriented face-on ($\theta_{\rm JN} = 0$) or face-off ($\theta_{\rm JN} = \pi$), $_{-2}Y_{21}(\theta_{\rm JN}, \phi)$ vanishes [169] despite a significant $h_{2,1}$ that results from the precession-induced mode mixing.

Therefore, in order to unambiguously measure precession with a dominant harmonic model, strong precession *and* an inclined system are required.

As a second example, we consider a waveform model that includes higher order coprecessing harmonics. In this case, even unfavorable orientations (i.e., face-on/off) allow the measurement of precession, despite the fact that along those directions, subdominant inertial harmonics [such as ($\ell, |m|$) = (2, 1) and (3,3)] are suppressed and cannot be measured. However, the dominant inertial harmonic reads,

$$h_{2,2} = \sum_{m' \in \pm\{1,2\}} h_{2,m'}^{\rm np} e^{i2\alpha} d_{m',2}^2(-\beta) e^{-im'\epsilon}, \quad (2.7)$$

which includes the coprecessing subdominant ($\ell = 2, |m| = 1$) harmonic. Here, measuring two different $\beta$ terms is already possible from the dominant inertial harmonic alone. Of course, $h_{2,1}^{\rm np}$ needs to be sufficiently strong. Because the leading-order amplitude of this harmonic is proportional to the mass difference of the two objects in the binary, sufficiently asymmetric systems are required.

Our two examples illustrate that in order to measure precession, two different $\beta$ terms can either be obtained for inclined systems that have strong enough subdominant inertial spherical harmonics. Or, if the signal (model) includes sufficiently strong subdominant coprecessing harmonics, then even a face-on or face-off binary can be identified as precessing despite detecting only the dominant inertial harmonic [77,127,169–171].

## III. WAVEFORM MODELS AND PARAMETER ESTIMATION

### A. Waveform model

The phenomenological family of waveforms accurately models the inspiral-merger-ringdown dynamics of a binary BH system. The post-Newtonian inspiral coefficients are interpolated to the intermediate and merger phases by fitting the unknown coefficients using numerical data. On the other hand, the ringdown coefficients are obtained employing the BH perturbation theory techniques. Though the early models were proposed mainly for searching GW signals buried in noisy data [172–175], we now have access to more accurate phenomenological waveform models for parameter estimation studies as well [77,155,176–179]. The state of the art gravitational waveform models include physical effects such as spin-induced orbital precession effects and subdominant harmonics other than the leading quadrupolar harmonic [164,180–184].

For our study, we mainly focus on IMRPhenomPv3HM waveform model [178] and use IMRPhenomPv3 waveform model to separate the effect of subdominant harmonics in measuring spin-induced orbital precession parameter ($\chi_p$). IMRPhenomPv3HM waveform model combines inputs from two other models, IMRPhenomHM [77,155] and IMRPhenomPv3 [179]. IMRPhenomHM is the first higher-harmonic waveform model for spinning binary BHs with spins aligned/antialigned to the orbital angular momentum axis, which incorporates subdominant harmonics $(\ell, |m|) = (3, 3), (4, 4), (2, 1), (3, 2), (4, 3)$ along with the dominant harmonic, $(\ell, |m|) = (2, 2)$. On the other hand, IMRPhenomPv3 accurately models the binary BH merger on a quasicircular orbit for generic spin orientations and accounts for the dominant harmonic in the coprecessing frame. One of the main improvements of IMRPhenomPv3 is the first phenomenological waveform model that accounts for two-spin precession [178,185,186].

A generic binary system on a quasicircular orbit is described mainly by a set of intrinsic parameters like the component masses, $m_i$ and the dimensionless spin vectors $\vec{S}_i$. To better understand the binary dynamics, one can define combinations of these parameters. For example, the mass asymmetry is measured by determining the ratios of component masses, $q = m_2/m_1$. More asymmetric the system as $q$ approaches to zero.

Two quantities capture the dominant spin effects associated with the orbiting binary system. The mass-weighted spin combination, known as the effective spin parameter [175], is defined by

$$\chi_{\text{eff}} = \frac{m_1\chi_1 + m_2\chi_2}{m_1 + m_2}, \quad (3.1)$$

and is argued to efficiently capture the dominant spin effects of a nonprecessing binary system where the spin angular momenta are aligned/antialigned to the orbital angular momentum axis. Here $\chi_1$ and $\chi_2$ are the components of the dimensionless spins along the direction defined by the orbital angular momentum (or perpendicular to the orbital plane), $\chi_i = \vec{S}_i \cdot \hat{L}/m_i^2$.

It is also possible that the spin angular momenta are not aligned/antialigned with the orbital angular momentum axis ($\hat{L}$). A spin misalignment induces orbital precession and is often quantified by the in-plane spin components and the angle between the spins. For an orbiting binary system, the magnitudes of in-plane spin components oscillate around the mean value as time evolves, and the angle between the in-plane spins changes monotonically for unequal mass systems. It has been shown that the dominant spin-precession effects can therefore be absorbed into an effective spin-precession parameter [187–189], $\chi_p$, which is the average value of the in-plane spin magnitudes over a large number of GW cycles. It is given by,

$$\chi_p := \frac{1}{A_1 m_1^2} \max(A_1 S_{1\perp}, A_2 S_{2\perp}), \quad (3.2)$$

where $A_1 = (2 + 3q/2)$ and $A_2 = (2 + 3/2q)$ are functions of component masses and $S_{i\perp} = |\hat{L} \times (\vec{S}_i \times \hat{L})|$ represents the in-plane spin component [160].

### B. Parameter estimation

Bayesian inference-based methods have been routinely employed to infer the properties characterizing the GW signal from binary coalescence. In this framework, the posterior distributions on each parameter $\theta$ is given by [190–195],

$$p(\theta|d, \mathcal{H}) = \frac{\mathcal{L}(d|\theta, \mathcal{H})\pi(\theta|\mathcal{H})}{\mathcal{Z}_\mathcal{H}}, \quad (3.3)$$

where $\mathcal{L}(d|\theta, \mathcal{H})$ is the likelihood function, $\pi(\theta|\mathcal{H})$ is the prior distribution, and $\mathcal{Z}_\mathcal{H}$ is the signal evidence assuming the hypothesis $\mathcal{H}$ being the model for the data. We use evidence as a normalization constant for this study.

We simulate binary BH signals assuming various source parameters describing the compact binary system at a fixed signal-to-noise ratio. For a binary system with total mass of 40 M$_\odot$ (detector frame), we consider three different mass ratios, $q = 1$ (equal mass system), $q = 0.28$ (GW190412 like system), and $q = 0.14$ (highly-asymmetric system). Although, indeed, the higher-harmonic contribution is a function of the total mass of the binary system, given the detector sensitivity, we restrict the analysis to fixed mass binaries. To understand the measurability of spin-induced orbital-precession effects, we consider a highly-precessing system with injected $\chi_p = 0.58$ and a slowly-precessing system with injected $\chi_p = 0.05$. This choice is made by fixing the individual spin magnitudes to be 0.6 and 0.3 and varying the spin vectors' angles accordingly. We fix the binary system's location so that each binary produces a three detector signal-to-noise ratio of 30. Further, we reanalyze the same set of signals fixing their locations closer so that the network signal-to-noise ratio is 60. Simulated binary signals contain both face-on and edge-on orientations, as the higher harmonic content in the signal increases from face-on to edge-on orientations. These software injections are made with zero-noise assumption and modeled using the IMRPhenomPv3HM waveform model for the majority of the cases. To compare the results, we also consider injections assuming the IMRPhenomPv3 waveform family. Table I summarizes the properties of the injected binary systems.

TABLE I. The properties of injected binary systems with total mass 40 $M_\odot$. The binary location is fixed so that it produces a three detector network signal-to-noise ratio of 30. For comparison, we also consider another set of injections with a signal-to-noise ratio of 60.

| Configuration | $q$ | $\chi_p$ | $\chi_{\text{eff}}$ | $\theta_{JN}$ | $D_L$ (Mpc) |
|---|---|---|---|---|---|
| A1 | 1.00 | 0.58 | 0.11 | 0 | 1032 |
| A2 | 1.00 | 0.05 | 0.44 | 0 | 1133 |
| B1 | 0.28 | 0.58 | 0.14 | 0 | 748 |
| B2 | 0.28 | 0.05 | 0.53 | 0 | 908 |
| C1 | 0.14 | 0.58 | 0.15 | 0 | 494 |
| C2 | 0.14 | 0.05 | 0.56 | 0 | 701 |
| D1 | 1.00 | 0.58 | 0.11 | $\pi/2$ | 349 |
| D2 | 1.00 | 0.05 | 0.45 | $\pi/2$ | 296 |
| E1 | 0.28 | 0.58 | 0.14 | $\pi/2$ | 380 |
| E2 | 0.28 | 0.05 | 0.53 | $\pi/2$ | 245 |
| F1 | 0.14 | 0.58 | 0.15 | $\pi/2$ | 343 |
| F2 | 0.14 | 0.05 | 0.56 | $\pi/2$ | 199 |
| G1 | 0.14 | 0.40 | 0.11 | $\pi/2$ | 243 |
| G2 | 0.14 | 0.30 | 0.11 | $\pi/2$ | 220 |

The prior on component masses is distributed uniformly over [3, 80] $M_\odot$. We further choose a uniform prior on the dimensionless spin magnitude, $|\chi_i| \leq 0.99$, and isotropic priors on the spin orientations. The prior on luminosity distance is uniform in [50,10000] Mpc. Reruns with different distance priors including the assumptions of uniform in comoving volume and uniform comoving four-volume confirm that our results and hence conclusions do not alter with respect to different distance prior choices. We consider a three detector network consisting of two LIGO detectors and the VIRGO detector. All the three detectors are kept at their respective design sensitivities [196–200].

Two essential quantities determining the strength of higher harmonics in the signal are the mass ratio and the inclination angle. See Sec. II for details. Therefore, higher harmonics in the signal help determining the orientation more accurately than dominant-harmonic signals could. To distinguish the effects of subdominant harmonics on intrinsic parameters from improved orientation measures, we analyze the simulated binary signals twice, once keeping the inclination fixed and once where it remains a free parameter in the recovery.

The marginalized posterior distributions on individual binary parameters are estimated using the open-source GW inference library parallel Bilby (pBilby) [201] assuming IMRPhenomPv3HM and IMRPhenomPv3 waveform models best describe the data. pBilby is a python based toolkit for GW data analysis where the stochastic sampling is performed using a dynamic nested sampling algorithm, dynesty [202].

## IV. MAIN FINDINGS

### A. The importance of including higher harmonics on $\chi_p$ measurements

We start with discussing the analysis of the injected signals A1–F2. See Table I for details. Our main interest is the recovered posterior distributions of the spin-precessing parameter, $\chi_p$. The results are shown in form of violin plots in Fig. 2. The top row shows the results for the strongly precessing cases, the bottom row for weakly-precessing cases. The panels on the left-hand side show the posterior distributions for face-on signals. The right-hand side panels show edge-on configurations. In Fig. 2, the IMRPhenomPv3HM waveform model is used for both injections and recoveries. For each injected signal, we show three distributions; the standard analysis where all parameters are unknown a priori is shown in green, if we fix the inclination angle to the value of the injected signal we obtain the orange distribution, and the light red distribution in the background is the prior distribution for $\chi_p$.

For an equal mass binary system, the $\chi_p$ distribution is less informative compared to an asymmetric system, independent of its orientations. We observe this feature for both highly-precessing and slowly-precessing cases. There is a significant improvement in $\chi_p$ measurements for the edge-on system compared to a face-on system and is more visible for higher mass ratio cases. According to Eq. (2.3), it is clear that many of the higher harmonics are nonvanishing for edge-on binaries hence improving the estimates.

The existence of subdominant harmonics most obviously helps to measure $\theta_{JN}$. To understand if the $\chi_p$ posteriors are affected by subdominant harmonics beyond correlations with a better-constrained $\theta_{JN}$, we repeat the analysis by fixing the inclination angle at the injected value. Comparing the green and orange violin plots in Fig. 2, we see that the posterior distributions on $\chi_p$ do not change significantly depending upon the inclination angle freedom in the analysis. Moreover, this introduces bias in the estimates and is more significant for slowly-precessing binaries.

Figure 3 shows the $\chi_p$ estimates obtained assuming IMRPhenomPv3 as the recovery model injecting both IMRPhenomPv3 (purple) and IMRPhenomPv3HM (blue) approximants. For the majority of the cases, there is a clear offset in the estimated $\chi_p$ distribution when we inject `IMRPhenomPv3HM` and recover using `IMRPhenomPv3`, pointing to the biases inducing from un-modeled effects such as the absence of higher harmonics in the approximant. We also see that the correlations between mass ratio and inclination angle improve as we include subdominant harmonics in the waveform but find no major impact on the $\chi_p$ estimates.

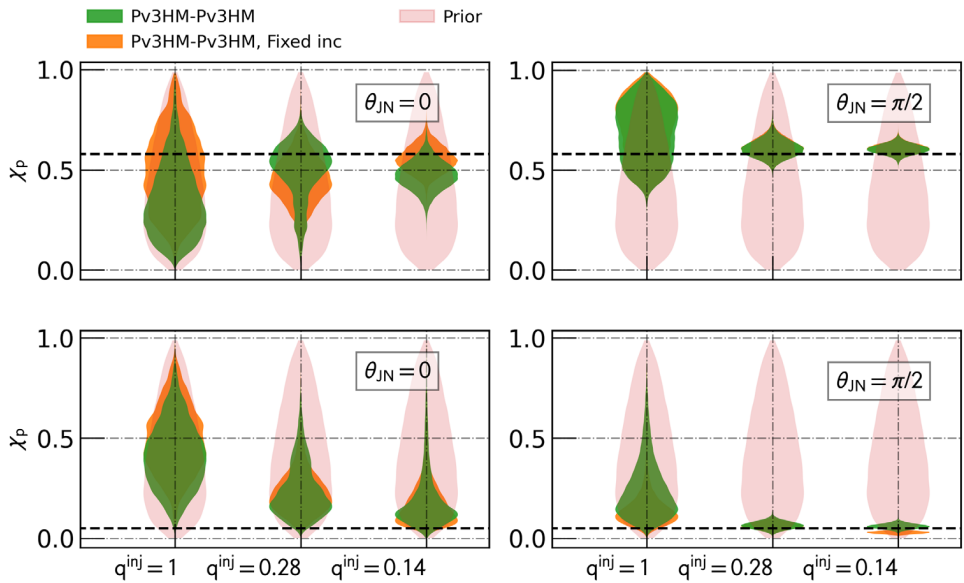

FIG. 2. The posterior distributions on the precessing spin parameter ($\chi_p$) for binary BHs injected of $\chi_p$ values 0.58 (top panel), and 0.05 (bottom panel) as indicated by the black dashed lines. Face-on and edge-on orientations for three mass ratios, $q = 1, 0.28, 0.14$ is considered. We use the `IMRPhenomPv3HM` waveform model for both injections and recoveries. Three plots represent the cases where inclination angle kept free in the recovery (green), inclination angle kept fixed (orange), and the prior distribution (light red).

Because we analyzed the same `IMRPhenomPv3HM` injections with models either including or not including higher co-precessing harmonics, we can calculate the ratio of evidence, i.e., the Bayes factors between both the model assumptions. Table II displays the logarithm (with base 10) $\log B_{\text{non}-\text{HM}}^{\text{HM}}$. Positive values indicate a preference for higher harmonics. As expected, binaries with large mass asymmetry and edge-on orientations show significant evidence for higher harmonics. Bayes factors obtained from equal-mass binaries are uninformative as higher co-precessing harmonics are suppressed. Recall that IMRPhenomPv3 does include the precession-induced (2,1) inertial harmonic, so the Bayes factors we calculate are not sensitive to whether or not this harmonic is present.

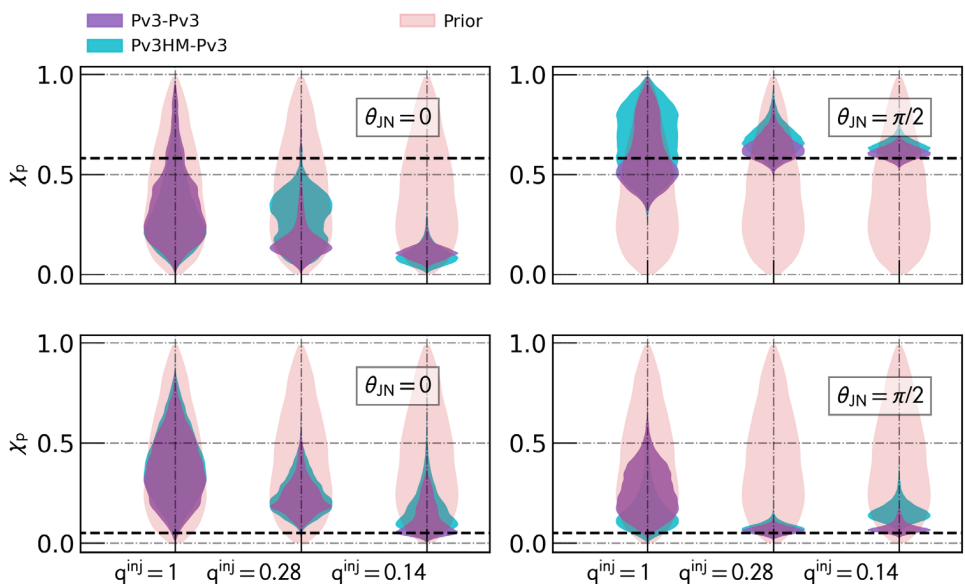

FIG. 3. The posterior distributions on the precessing spin parameter ($\chi_p$) for binary BHs of injected $\chi_p$ values 0.58 (top panel) and 0.05 (bottom panel). We consider three mass ratios for both face-on and edge-on inclinations. Purple violin plots show the estimates when `IMRPhenomPv3` is used as both injection and recovery model, whereas the blue compares the performance of `IMRPhenomPv3HM` over `IMRPhenomPv3` as the recovery model. The black dashed lines indicate the injected value.

TABLE II. Bayes factors comparing the two hypotheses, signal containing higher harmonics and the signal contains only the dominant harmonic. We use IMRPhenomPv3HM and IMRPhenomPv3 waveform models respectively to represent the signals with and without higher modes present.

| Configuration | $\chi_p$ | $q$ | $\theta_{JN}$ | $\log B^{HM}_{non-HM}$ |
|---|---|---|---|---|
| A1 | 0.58 | 1 | 0 | −0.46 |
| A2 | 0.05 | 1 | 0 | −0.14 |
| B1 | 0.58 | 0.28 | 0 | 2.55 |
| B2 | 0.05 | 0.28 | 0 | −1.75 |
| C1 | 0.58 | 0.14 | 0 | 13.46 |
| C2 | 0.05 | 0.14 | 0 | 23.84 |
| D1 | 0.58 | 1 | $\pi/2$ | 0.24 |
| D2 | 0.05 | 1 | $\pi/2$ | −0.19 |
| E1 | 0.58 | 0.28 | $\pi/2$ | 36.12 |
| E2 | 0.05 | 0.28 | $\pi/2$ | 33.8 |
| F1 | 0.58 | 0.14 | $\pi/2$ | 42.04 |
| F2 | 0.05 | 0.14 | $\pi/2$ | 59.6 |

The slowly-precessing face-on binary with $q = 0.28$ (B2) is a curious case. Here, the Bayes factor favors the dominant harmonic model. We speculate that higher co-precession harmonics are significant for many configurations in the prior parameter space at this mass ratio. However, the injected signal is chosen at a specific point where the inclination and the weak precession suppress higher harmonics. Therefore, the lack of measurable harmonics favors the dominant-mode model that does not receive the same punishment in the likelihood when inclined or more strongly precessing templates are compared to the data. Of course, our prior knowledge of the underlying theory is that higher harmonics are part of real signals, so one must not conclude that the dominant-mode model yields a more accurate measurement, despite the Bayes factor favoring it.

We note that we reanalyzed some of the cases listed in Table I with the NRSur7dq4 model [203] to verify that the main findings remain unaltered across different waveform families. NRSur7dq4 is a numerical-relativity surrogate model that describes the inspiral-merger-ringdown dynamics of precessing BBHs and proved to be accurate for binaries with $M > 60 M_\odot$ and inverse mass ratio $1/q > 4$ (see Fig. 9 of [203]). We find good agreement between the results despite the differences in the waveform generation and calibrated parameter region. Specially, we find that the edge-on binaries allow $\chi_p$ measurements with good accuracy for all mass ratios. In contrast, face-on binaries with equal masses show biased $\chi_p$ measurements due to the lack of subdominant harmonics.

### B. Precession measurements with the dominant harmonic model and the role of prior distribution

We now focus on the possible biases and constraints on $\chi_p$ when the dominant harmonic model IMRPhenomPv3 is employed both as the injected signal and as the template model. While this constitutes a scenario with over-simplified signals, it allows us to study the fundamental limitations of dominant harmonic models separately from systematic differences between signal and template models. The resulting posterior distributions on $\chi_p$ are shown as violet violin plots in Fig. 3.

For the face-on case ($\theta_{JN} = 0$), the dominant harmonic model fails to distinguish a highly-precessing system from a slowly-precessing system. The $\chi_p$ estimates are biased towards lower values as we increase the mass asymmetry for a highly-precessing system. Whereas, for slowly-precessing systems, we find $\chi_p$ posteriors to be overestimated.

We use a vanishing noise realization in the data we analyze, and there is no difference between the injected model and the template family. So, where does the bias come from? As the component spins are comparatively unconstrained, the posterior information on $\chi_p$ must be derived from prior assumptions and constraints on other parameters. As evident from Eq. (3.2), $\chi_p$ is correlated with $q$. Even though we start with uninformative priors for spins and individual masses, as soon as the mass ratio is constrained by the data, we will infer probable values for $\chi_p$ that may incorporate little or no information about the spins themselves. They are dominated by the $\chi_p$ prior restricted to the measured mass-ratio range.

To illustrate this effect, we derive explicit expressions for the prior distribution of $\chi_p$ for fixed values of $q$. As mentioned previously, our uninformative prior assumes isotropic spin orientations and uniformly distributed spin magnitudes. These assumptions imply that each BH's in-plane spin follows the prior distribution,

$$\pi_\perp(\chi_\perp) = \arccos(\chi_\perp), \quad (4.1)$$

as can be seen through straightforward coordinate tranformations. According to Eq. (3.2), $\chi_p$ takes the value of either the more massive BH's in-plane spin $\chi_{1\perp}$, or the in-plane spin parameter $\chi_{2\perp}$ multiplied with $\kappa = q^2 A_2/A_1$—whichever value is greater. To calculate the prior distribution of the maximum of these two numbers, we first need the cumulative distribution,

$$F(X) = P(\chi_\perp < X) = \int_0^X \pi_\perp(\chi_\perp) d\chi_\perp$$
$$= 1 - \sqrt{1 - X^2} + X \arccos(X). \quad (4.2)$$

From the product of the two cumulative distributions for $\chi_{1\perp}$ and $\kappa\chi_{2\perp}$, we finally derive an expression for the $\chi_p$ prior,

$$\pi(\chi_p) = \frac{d}{d\chi_p}[P(\chi_{1\perp} < \chi_p) \cdot P(\kappa\chi_{2\perp} < \chi_p)]$$

$$= \begin{cases} \pi_\perp(\chi_p)F(\chi_p/\kappa) + \frac{F(\chi_p)}{\kappa}\pi_\perp(\chi_p/\kappa), & \chi_p < \kappa \\ \pi_\perp(\chi_p), & \chi_p \geq \kappa \end{cases}. \quad (4.3)$$

In Fig. 4, we show the result of our analytical calculation together with a numerical realization of the $\chi_p$ prior for different mass ratios $q = 1, 0.2, 0.01$. For $q = 1$, one has $\kappa = 1$ and $\pi(\chi_p) = 2F(\chi_p)\pi_\perp(\chi_p)$, which is a curve that peaks at $\chi_p \approx 0.58$ and falls off gradually to either side of the peak. Conversely, for very small $q$, the case $\chi_p < \kappa$ in Eq. (4.3) becomes negligible, and the $\chi_p$ prior follows the prior of the primary BH's spin, $\pi(\chi_p) = \pi_\perp(\chi_p) = \arccos(\chi_p)$. For moderate mass ratios, the $\chi_p$ prior distribution has a characteristic peak at low $\chi_p$ values, caused by the fact that even large in-plane spins on the secondary BH may only lead to small values of $\chi_p$.

We find the same trend in the posteriors of the dominant harmonic model and face-on orientations: As the mass ratio is constrained towards small values, the $\chi_p$ probability shifts towards small values following the prior. To visualise the actual constraints derived from the data, we show the scatter plots for $\chi_{\rm eff}$ and $q$ in Fig. 5 for a highly-precessing face-on binary system employing the IMRPhenomPv3 model. The different colors indicate three different mass ratios as marked by the plus mark, and the orange scatter plots represent the prior distribution. While the injected value can be located at the edge of the two-dimensional posterior region, we see that the posterior ranges on $\chi_{\rm eff}$ and $q$ are tightly clustered around the true value. Therefore, even though the $\chi_p$ posterior follows the prior distribution, the $\chi_{\rm eff} - q$ plane is well constrained compared to the prior distribution, especially for highly-asymmetric binaries.

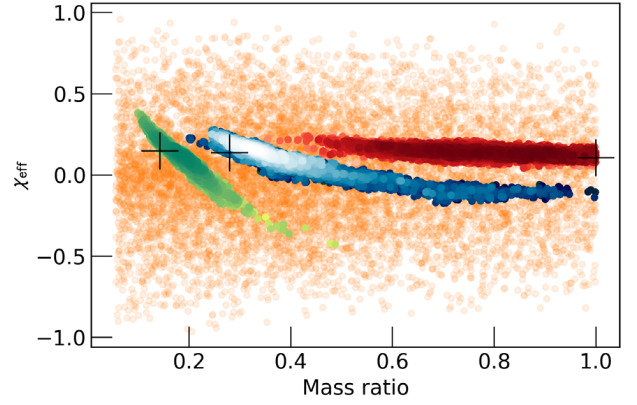

FIG. 5. Effective spin parameter $\chi_{\rm eff}$ and mass ratio scatter plots for a highly-precessing face-on binary system for three mass ratios, $q = 1$ (red), $q = 0.28$ (blue), and $q = 0.14$ (grey). The orange scatter plots in the background are the prior distribution. IMRPhenomPv3 waveform model is used for both injections and recoveries. The plus mark shows the injected values for each case.

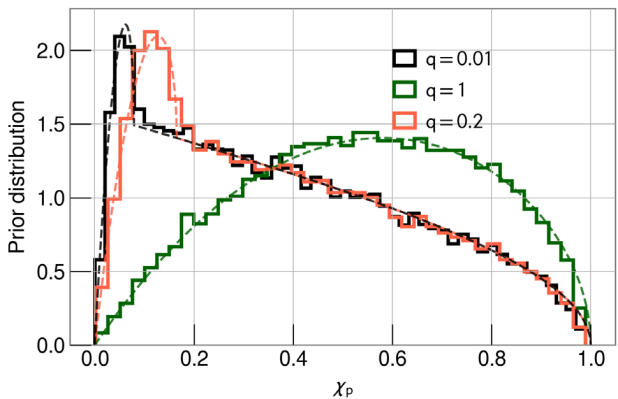

FIG. 4. The prior distribution on $\chi_p$ fixing the mass ratios to be $q = 1, 0.2, 0.01$. As we increase the inverse mass ratio ($1/q$) the prior distribution on $\chi_p$ shifts towards low values. The structure that appears at low $\chi_p$ values for asymmetric binaries is because in those systems, even large in-plane spins of the less massive BH may only lead to a small value of $\chi_p$. See (4.3) for the analytical expressions.

For edge-on inclinations, the IMRPhenomPv3 waveform model can measure the spin-precession parameter much more reliably (see the right panel in Fig. 3), though the estimates are less tight and accurate than those in Fig. 2. For strongly inclined sources, the $\chi_p$ posterior is not prior dominated. The data include significant contributions from the intertial $(\ell, |m|) = (2, 2)$ and $(2,1)$ modes when the system is precessing, which gives sufficient information to constrain $\chi_p$.

Overall, when we compare the performances of the two waveform models for measuring $\chi_p$, it is clear that both models can provide evidence for precession, certainly for unequal mass binaries with edge-on orientations. However, the posterior distribution tends to shrink more towards the correct value (injected value) when we apply the IMRPhenomPv3HM model, indicating the importance of using subdominant harmonics in the waveform, as we discussed in Sec. II.

### C. The effect of signal-to-noise ratio on $\chi_p$ measurements

To better understand the possible improvement in measuring $\chi_p$ for a binary system observed with a larger signal-to-noise ratio in the three detector network, we compare the SNR $= 30$ case with another set of estimates by doubling the signal-to-noise ratio, employing IMRPhenomPv3HM as the recovery model. Figure 6 shows the violin plots on one-dimensional marginalized posterior distributions on $\chi_p$ by varying the mass ratios from $q = 1, 0.28, 0.14$ considering face-on (left panel) and edge-on (right panel) orientations. The top panel in the figure shows the case for an injected $\chi_p$ of 0.58 whereas the bottom panel is that of a slowly-precessing system with injected $\chi_p$ of 0.05.

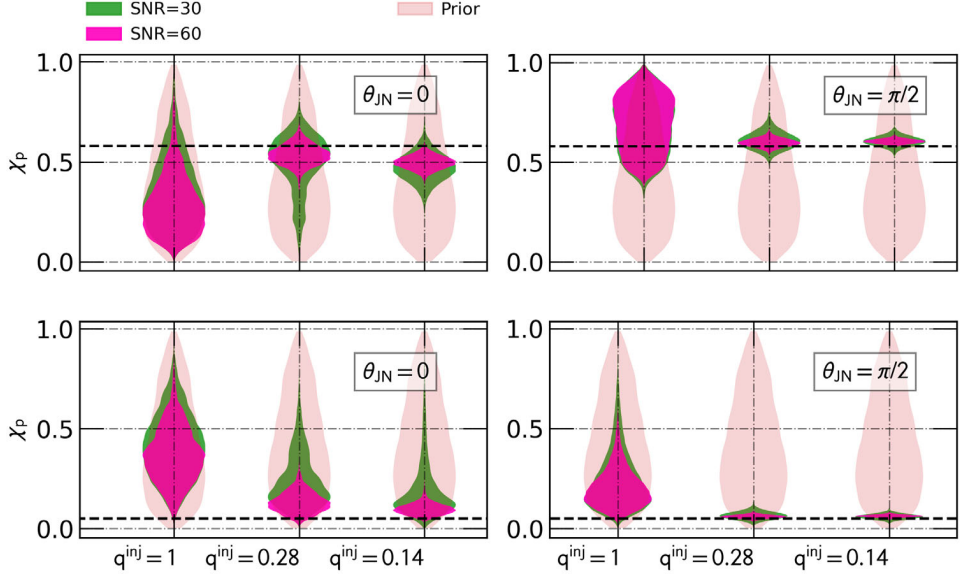

FIG. 6. Posterior distributions on $\chi_p$ for binaries with three different mass ratios, $q = 1, 0.28, 0.14$ considering both face-on and edge-on orientations. The top and bottom panels respectively show the cases for injected $\chi_p$ values of 0.58 and 0.05, respectively. The green plots with the SNR = 30 assumption are over-plotted with the SNR = 60 cases (pink) and the injections as black dash lines.

We see a significant improvement in the estimates as we double the SNR for asymmetric binaries with face-on or edge-on orientations. However, for a face-on equal mass binary system increasing the SNR does not help break correlations to result in better measurements of $\chi_p$. That means for a face-on equal mass binary system, our knowledge on $\chi_p$ does not improve much regardless of the injected $\chi_p$ value.

The difference in the $\chi_p$ posterior distributions for slowly-precessing and highly-precessing cases is solely coming from the fact that we analyze two different injections.

To highlight this, we show the effective spin-parameter ($\chi_{\rm eff}$) and mass ratio ($q$) scatter plots in Fig. 7, for a face-on binary system with injected $\chi_p$ values of 0.58 (top panel) and 0.05 (bottom panel). In both the cases, as the red (SNR = 60) and blue (SNR = 30) scatter plots indicate, the $\chi_{\rm eff}$ and $q$ estimates improve as the SNR doubles but not enough to improve the $\chi_p$ estimates. Additionally, see that the properties of these measurements have notable differences from slowly-precessing to highly-precessing binaries.

From this SNR comparison study, we emphasize that the SNR has a visible role in accurate $\chi_p$ measurements for the majority of the case we analyze here. Despite this, the effect of SNR is negligible for equal-mass binaries with face-on or edge-on orientations.

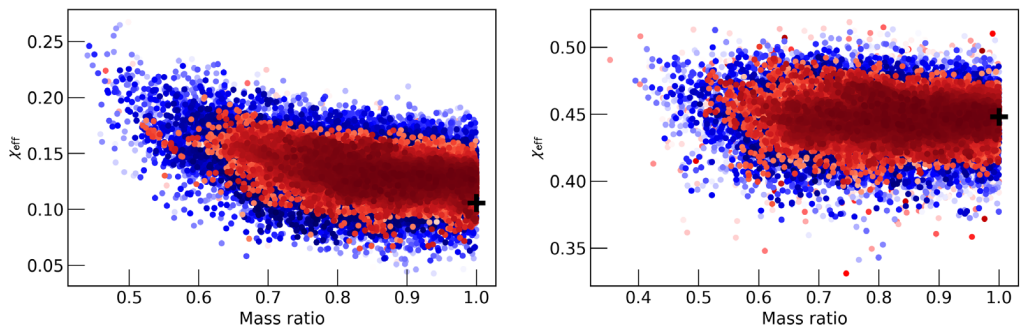

FIG. 7. Effective spin parameter $\chi_{\rm eff}$ and mass ratio scatter plots for a highly-precessing face-on binary system (top panel) and a slowly-precessing system (bottom panel). Red and blue scatter points correspond to SNR = 60 and SNR = 30 cases, respectively. IMRPhenomPv3HM waveform model is employed for both injections and recoveries. The plus mark shows the injected values for each case.

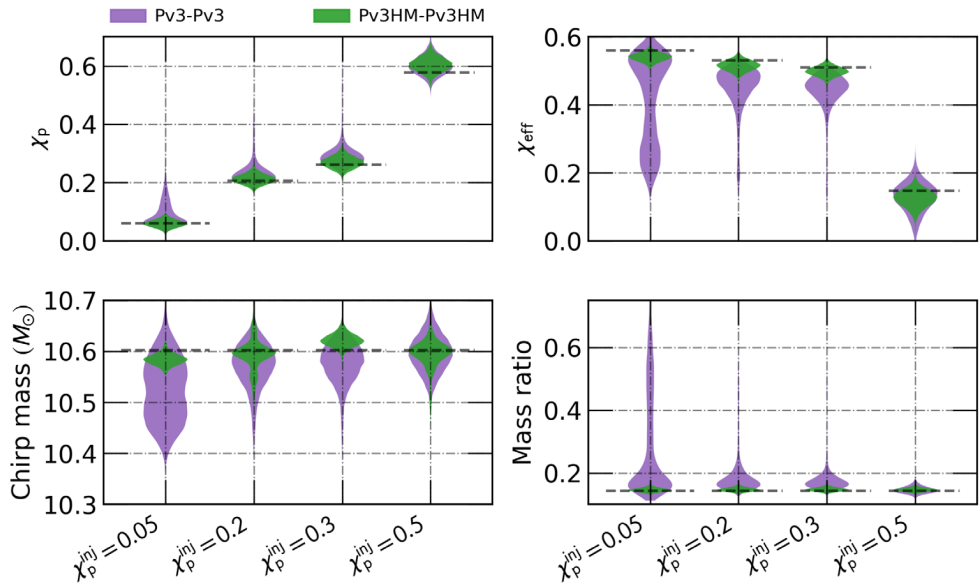

FIG. 8. Posterior distributions on the chirp mass, mass ratio, $\chi_p$ and $\chi_{\rm eff}$ for different values of injected spin-precession parameter $\chi_p = 0.05, 0.2, 0.3, 0.5$ for an edge-on binary with component masses $(35, 5)\ M_\odot$.

### D. Measurement accuracies for highly-asymmetric edge-on binaries with varying spin-precession effects

From our analysis, we see that both waveform models perform reasonably well for measuring spin-precession in the case of edge-on asymmetric binaries (see Fig. 2 and Fig. 3). As predictable from the discussion in Sec. II, in this case, many subdominant harmonics contribute to IMRPhenomPv3HM. At the same time, the extra harmonic arising from the precession-induced mode-mixing contributes in the case of IMRPhenomPv3.

To compare the performance of the two waveform models beyond the $\chi_p$ measurement, we show the chirp mass, mass ratio ($q$), effective spin parameter ($\chi_{\rm eff}$), and effective spin-precession parameter estimations ($\chi_p$) for a binary system with mass ratio 0.14 and edge-on orientation in Fig. 8. Notably, we vary the injected $\chi_p$ values from 0.05, 0.2, 0.3 and 0.5 to examine the importance of spin-precession in such cases.

Along with $\chi_p$, estimations of all other parameters improve when we model the signal using IMRPhenomPv3HM as the green violin plots indicate. The dominant harmonic model requires highly-processing systems to provide better measurements of chirp mass, mass ratio, and the effective spin parameter. This is because for an edge-on binary with significant mass asymmetry, the spin precession helps breaking the degeneracy between the mass-spin parameters [171] resulting in the most accurate set of estimates, especially for IMRPhenomPv3. When higher harmonics are included, additional mass-ratio information is already present from the strength of the harmonics. Therefore, the IMRPhenomPv3HM analysis does not require strongly-processing sources for an accurate measurement of the mass and spin parameters.

## V. SUMMARY

The recent catalog of binary signals released by the LIGO-VIRGO Collaboration contains binary signals with varying properties. We expect to see many more such events in the future, including highly-processing binaries with mass asymmetries and edge-on orientations. By performing parameter estimation analysis on simulated BBH signals, with the IMRPhenomPv3HM waveform model, we show that higher harmonics permit us to infer the presence of precession even for face-on binaries with mass-asymmetry. On the other hand, the dominant harmonic model fails to extract enough information on the spin-precession for moderate signal-to-noise ratio signals. With this, we emphasize the importance of using waveform models with higher harmonics and spin-precession effects for parameter inference of binaries with face-on or edge-on orientations. However, even with the IMRPhenomPv3HM waveform model, it is challenging to infer accurate information on the spin-precession if the binaries are equal mass and have face-on orientations. Furthermore, the increased network signal-to-noise ratio helps improve the measurement accuracy for unequal mass face-on systems and all systems with edge-on orientations, except for face-on equal mass binaries with large spin-precession.

## ACKNOWLEDGMENTS

This work has supported by the Max Planck Society's Independent Research Group Grant. Computations were carried out on the Holodeck cluster of the Max Planck Independent Research Group "Binary Merger Observations and Numerical Relativity" and the Max

Planck Computing and Data Facility computing cluster Cobra. We thank Eleanor Hamilton for carefully reading the manuscript and providing us useful comments and suggestions. We are grateful to Mark Hannam and Sebastian Khan for useful discussions. We thank the anonymous referee for the useful comments and suggestions on the manuscript. This document has LIGO preprint number P2100196.